# Edge Deployment of Small Language Models, a comprehensive comparison of CPU, GPU and NPU backends

Pablo Prieto, Pablo Abad

***Abstract*— In the era of the Internet of Things (IoT), edge computing has become a key paradigm, enabling data to be processed at the point of generation rather than relying on remote cloud servers. This approach supports faster decision-making, reduces bandwidth usage, and enhances privacy. However, edge devices typically operate under strict constraints on processing power, memory, and energy consumption, making them unsuitable for large language models (LLMs), which continue to grow in size and computational demands. Recently, a growing ecosystem of Small Language Models (SLMs) has been introduced, offering lightweight alternatives that bring AI inference to resource-constrained environments by significantly reducing computational cost while remaining suitable for specialization and customization.**

**These resource limitations impose strict requirements on the choice of hardware platform, and it is often unclear which option provides the best balance between performance and efficiency. To address this challenge, this study evaluates the inference performance and energy efficiency of commercial CPUs (Intel and ARM), GPUs (NVIDIA), and NPUs (RaiderChip) for running SLMs. GPUs, known for their parallel processing capabilities, are compared against commercial NPUs and recent multi-core CPUs. While NPUs leverage custom hardware designs to optimize computation, modern CPUs increasingly incorporate dedicated features targeting language model workloads. Using a common execution framework and a suite of state-of-the-art SLMs, we analyze the maximum achievable performance, as well as processing and energy efficiency across commercial solutions available for each platform.**

**The results indicate that specialized backends outperform general-purpose CPUs, with NPUs achieving the highest performance by a wide margin. Bandwidth normalization proves essential for fair cross-architecture comparisons. Although low-power ARM processors deliver competitive results when energy usage is considered, metrics that combine performance and power (such as EDP) again highlight NPUs as the dominant architecture. This shows that designs optimized for both efficiency and performance offer a clear advantage for Edge workloads.**

All authors are with the Computer Engineering research group, Universidad de Cantabria, 39005 Santander, Cantabria (e-mail:{prietop,abadp}@unican.es)

## I. Introduction

THE last decade has been a blast for artificial intelligence, with large language models (LLMs) stunning everyone with their capabilities and powering everything from chatbots to code assistants. Despite their success, the massive size and computational demands of LLMs makes them impractical for many real-world scenarios [1][2][3]. Deploying such models on edge devices, such as mobile phones or IoT nodes, is significantly challenging due to strict constraints on memory, processing power and energy consumption.

To address this limitation, Small Language Models (SLMs) have recently emerged as lightweight alternatives, designed to bring powerful AI inference to resource-constrained environments. These models typically contain up to a few billion parameters, orders of magnitude fewer than the hundreds of billions found in LLMs, resulting in lower memory footprints and computational requirements. Their efficiency makes them particularly attractive for edge devices and scenarios where privacy demands offline inference [4]. Moreover, SLMs have been recently highlighted as especially well-suited for the emerging field of agentic AI [5], where most agents do not require the full generalist capabilities of LLMs. Their smaller scale facilitates fine-tuning and specialization for narrow tasks.

The creation of leaner models often relies on compression techniques [6] such as pruning, quantization, low-rank factorization and distillation, trying to preserve as much accuracy as possible. The growing practical value of SLMs has fueled rapid progress in this area, with a continuous stream of new models being proposed to balance accuracy, performance and efficiency [7][8][9][10]. This rapid model evolution, however, raises new challenges for hardware design. While GPUs have long been the platform of choice for running language models, the demand for efficient AI inference has prompted broader innovation. On one hand, general-purpose processors are increasingly equipped with features to accelerate the algebraic operations underpinning neural networks [11]. On the other hand, specialized architectures such as Neural Processing Units (NPUs), often built on FPGA or ASIC devices, directly target inference workloads by optimizing core arithmetic operations [12][13][14][15]. In this heterogeneous and fast-changing landscape, it remains difficult to maintain a clear understanding of the strengths and limitations of different execution platforms. This paper aims to address this gap by systematically evaluating four commercial hardware platforms (x86-64 and ARM CPUs, NVIDIA GPU, and RaiderChip NPU) suitable for running SLMs in edge environments. Using a diverse set of state-of-the-art models in both original and quantized forms, we assess performance, energy consumption and resource utilization. Our goal is to provide practical insights that support software engineers and system designers in selecting the most effective hardware platform for SLM inference at the edge.

## II. Background

The focus of this work is the inference process in transformer-based language models, since this is the stage directly executed on edge backends. While lighter than training, as it only requires forward propagation, it remains computationally demanding because it involves large matrix multiplications in the attention mechanism and feed-forward layers, complemented by normalization, activation and embedding operations.

Efficient inference depends not only on suitable hardware but also on the software stack that bridges model availability and execution. Repositories such as Hugging Face ensure standardized access to pre-trained models, while local inference frameworks provide optimized implementations for specific backends. In addition, quantization techniques play a central role in enabling SLMs to fit within the resource budgets of edge devices. As outlined in Figure 1, the following subsections describe the software stack employed in this work and the quantization strategy adopted in our evaluation.

### *A. SLM Software Stack*

The Hugging Face Hub, the most popular repository for language models, employs a tensor-only model format designed for safety and efficiency. This format, known as *safetensors*, stores only tensor weights along with a lightweight JSON header that specifies metadata such as architecture details and tokenizer information. By avoiding the execution of arbitrary code, it eliminates common security risks. In addition, libraries can memory-map (mmap) tensors directly from disk, enabling fast model loading with minimal memory spikes. Primarily used for model distribution, local inference frameworks often convert models into their own formats derived from *safetensors*.

One of the most widely adopted formats for local execution is GGUF, a tensor-only format optimized for quantized models. GGUF supports low-bit weight representations along with scaling factors and zero-points, making it well suited for compact model storage and efficient inference. It is integrated into the Hugging Face Hub and supported by inference frameworks such as llama.cpp, GPT4All, and Ollama.

Llama.cpp [16] is an open-source inference framework originally designed for LLMs based on Meta's LLaMA architecture[17][18]. Implemented in C++, it is built on top of the GGML tensor library [19], which provides optimized matrix operations across diverse hardware. A key feature of llama.cpp is its broad support for quantization, offering formats from Q2 to Q8. These enable LLM execution on consumer-grade hardware with limited memory while preserving acceptable performance. The framework also offers broad backend compatibility, including CPUs with AVX/AVX2/AVX512 instruction sets and GPUs through CUDA, Metal, Vulkan, and OpenCL. This combination of quantization flexibility and hardware support makes llama.cpp a general-purpose framework for LLM deployment across heterogeneous platforms.

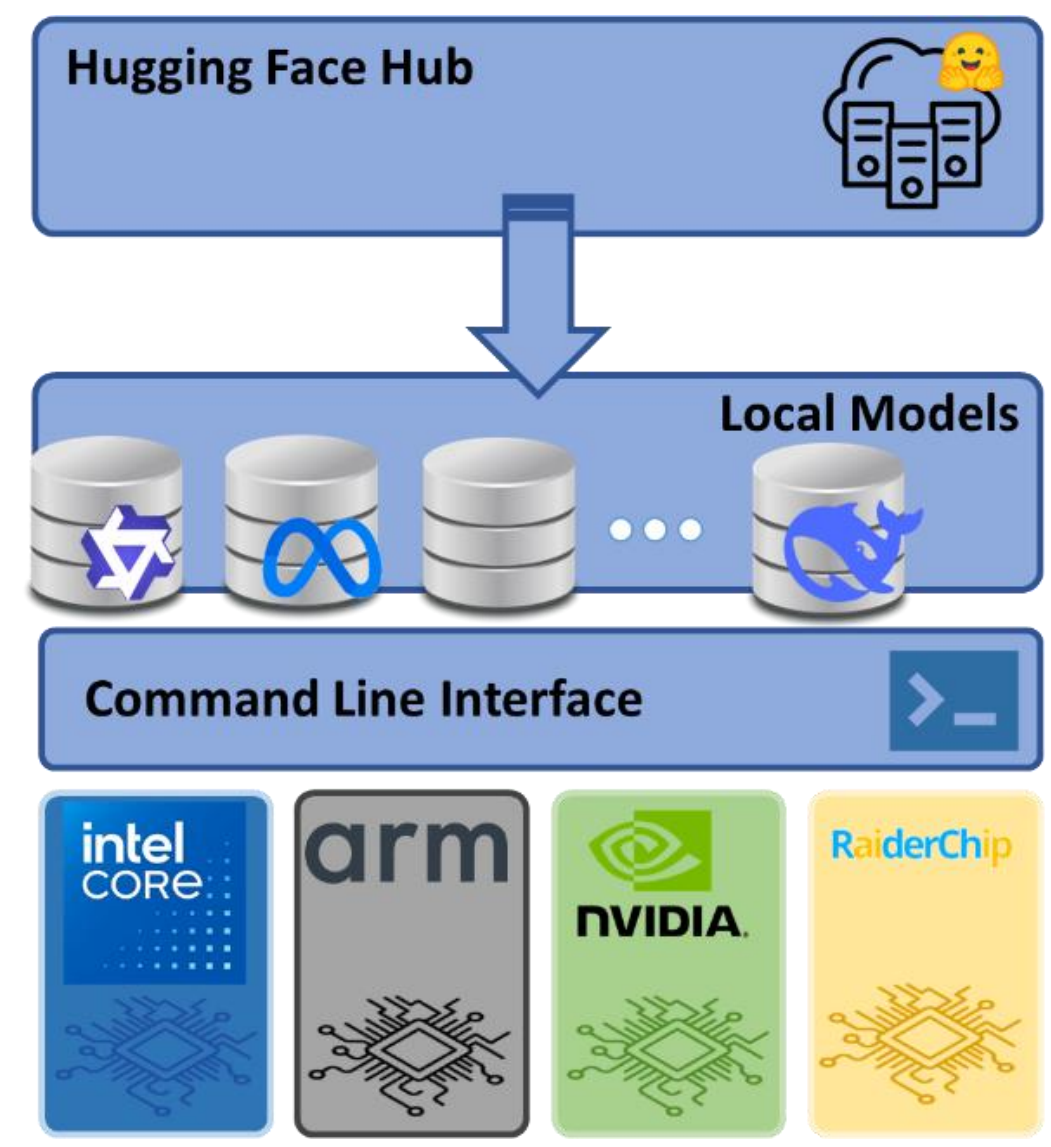

Figure 1. Evaluation framework, software stack.

The selection of backends evaluated in the next section was guided by two key requirements for the hardware–software stack. First, the stack must support models distributed through the Hugging Face Hub in their original *safetensors* format. Second, it must be capable of handling both quantized models and their native floating-point (f32 or f16) versions, ensuring that accuracy can be preserved when full-precision weights are required. To meet these requirements, all models were obtained from the Hugging Face Hub, converted into the appropriate runtime format (*GGUF* via *llama.cpp* for CPU and GPU, or a custom format for the NPU), and executed through an optimized low-level framework (*llama.cpp* for CPU and GPU, and a custom framework with an equivalent CLI interface for NPU[1])

### *B. Quantization.*

Quantization is one of the most popular techniques for reducing the size and computational requirements of models without significantly compromising their accuracy [20]. There are multiple quantization mechanisms, typically classified according to the stage of the model lifecycle in which they are applied: training and inference (Quantization Aware Training) versus inference only (Post-Training Quantization). Additionally, quantization can be applied to different components of the process, such as weights, activations, or the key-value (KV) cache.

[1] The conversion process, as well as tasks such as weight uploading are fully automated and transparent to the user, being integrated in the RaiderChip NPU interface.

One of the most widely adopted techniques, due to its simplicity and results, is weight-only quantization. In this case, each FP32/F16 value from the original model is represented using a low-precision data type, typically ranging from 8 bits down to 2 bits in more extreme cases. This significantly reduces the model's memory footprint and enables more efficient use of available bandwidth. From a computational standpoint, both quantization and dequantization (the process of converting a reduced-precision value back to one as close as possible to the original) are straightforward and require very few operations. This makes models highly versatile, as they can be stored in reduced precision while still operating at full precision. As computing resources can be a limiting factor in edge environments, quantized models are usually deployed on edge devices to overcome technical challenges. For this reason, all evaluated models are used in two versions: the original one with F32 or F16-weight precision and a quantized version Q4K.

## III. Methodology

### *A. Models.*

This subsection presents the Small Language Models (SLMs) selected for evaluation. All models are publicly available through the Hugging Face Hub. Our selection prioritizes diversity in architecture, training methodology, and model scale, while adhering to a strict upper limit of 14 billion parameters to ensure compatibility across all evaluation backends. In total, 11 models were selected, as summarized in Table 1. The models, organized into four groups according to their family, are briefly described next.

TABLE 1: SLM MODELS UNDER EVALUATION.

| Model (Label) | Context | Hidden layers | Att. Heads | Hidden Size |
|---|---|---|---|---|
| Qwen3-8B (Q3_8B) | 32K | 36 | 32 | 4096 |
| Qwen3-1.7B (Q3_1.7B) | 32K | 28 | 16 | 2048 |
| Qwen3-0.6B (Q3_0.6B) | 32K | 28 | 16 | 1024 |
| Qwen2.5-Coder-1.5B (Q2.5_1.5B) | 32K | 28 | 12 | 1536 |
| Llama-3.1-8B-Instruct (L3.1_8B) | 8K | 32 | 32 | 4096 |
| Llama-3.2-3B-Instruct (L3.2_3B) | 8K | 28 | 24 | 3072 |
| Llama-3.2-1B-Instruct (L3.2_1B) | 8K | 16 | 32 | 2048 |
| DSeek-R1-0528-Qwen3-8B (DSQ3_8B) | 32K | 36 | 32 | 4096 |
| DSeek-R1-Distill-Qwen-14B (DSQ_14B) | 32K | 48 | 40 | 5120 |
| DSeek-R1-Distill-Llama-8B (DSL3_8B) | 32K | 32 | 32 | 4096 |
| Falcon3-1B-Instruct (F3_1B) | 8K | 18 | 8 | 2048 |

Four models from the Qwen series (version 3 [21] and 2.5 [22]) were selected to represent the latest developments in Alibaba's open-source foundation models:

- Qwen3-8B: A dense transformer model with ~8 billion parameters, featuring an updated tokenizer, grouped-query attention (GQA), and support for multi-modal extensions.
- Qwen3-1.7B: A smaller variant (~1.7B parameters) with architectural consistency to the 8B model, useful for assessing scale-dependent behaviors.
- Qwen3-0.6B: The smallest model in the series (~0.6B parameters), offering insight into the capabilities of ultra-compact transformers.
- Qwen2.5-Coder-1.5B: A 1.5B parameter code-focused variant from the Qwen 2.5 line, fine-tuned for programming-related tasks with an extended tokenizer and code-aware training corpus.

Four models from the Meta-Llama [17][18] series are included to capture variations across model versions, scales, and instruction tuning strategies:

- Meta-Llama-3.2-3B-Instruct: A 3B parameter instruction-tuned model from Meta's LLaMA 3.2 release, optimized for general-purpose tasks.
- Meta-Llama-3.1-8B-Instruct: An 8B instruction-tuned model reflecting the 3.1 version of the LLaMA 3 family, offering enhanced performance in reasoning and language tasks.
- Meta-Llama-3.2-1B-Instruct: A compact 1B parameter model for efficiency-critical applications, instruction-tuned for improved alignment.
- Meta-Llama-2-7b-chat-hf: A 7B parameter chat-optimized model from the LLaMA 2 generation, instruction-tuned for conversational tasks and provided in Hugging Face format.

Three models are derived from the DeepSeek-R1 family [23], which integrates elements from LLaMA and Qwen. These models are designed to balance performance and efficiency via distillation:

- DeepSeek-R1-0528-Qwen3-8B: The latest 8B release as of May 2024, incorporating improvements in inference speed, reasoning, and factual accuracy.
- DeepSeek-R1-Distill-Qwen-14B: A distilled variant of a 14B Qwen model, reduced in size for small-model deployment while preserving key capabilities.
- DeepSeek-R1-Distill-Llama-8B: A distilled version of an 8B LLaMA-based model, maintaining core performance while offering a more compact footprint.

Finally, a 1B parameter instruction-tuned model from the Falcon 3 series [24] is included, designed for task completion and low-latency applications, based on a transformer architecture with multi-query attention.

### *B. Execution Backends.*

This section describes the four hardware platforms evaluated in this study, emphasizing those architectural aspects most relevant to small language model (SLM) inference. All platforms support execution of full-precision models, as distributed through Hugging Face Hub (F16), as well as their Q4K quantized counterparts.

For GPU-based edge inference, the selected platform is the NVIDIA Jetson Orin Nano Super module with an Ampere-architecture GPU comprising 1024 CUDA cores and 32 third-generation Tensor Cores. The Tensor Cores accelerate matrix-intensive operations typical of language models and support multiple numerical precisions, including INT8, which is critical for efficient SLM inference. The module includes 8 GB of 128-bit LPDDR5 memory, providing a peak theoretical bandwidth of 102 GB/s. Because it is a popular platform for edge computing, we will use this backend as a reference device in our evaluation process.

The first general-purpose platform is a desktop-class x86-64

CPU. Even though it is not a system specifically designed for edge computing, it represents a more modest configuration compared to a server system, closer to the configuration of edge platforms, and represents the ubiquity of x86 desktop class CPUs. The platform has an intel i5-12400 processor with 6 cores supporting AVX2 instruction set. It has two memory channels providing a theoretical memory bandwidth of approximately 76.7GB/s. To enrich the evaluation process, we added a second CPU-type backend using the ARM processor included in the NVIDIA Jetson platform. This architecture has six Cortex-A78AE cores with v8.2 64 bits computing capabilities, including up to 128 bits vector instructions. The ARM processor has access to 6GB of LPDDR5 memory with 102GB/s available bandwidth.

The RaiderChip NPU AI accelerator is fully implemented as Programmable Logic inside an AMD Versal FPGA device (VH1582) which includes 32 GB of HBM2e memory interconnected at 102 GB/s, in order to match the memory bandwidth of the NVIDIA Jetson platform. The NPU contains eight Matrix Multiplication Engines operating thousands of floating-point operations per cycle (from Q4 to FP32 natively), complemented by a Vector Engine (16 ops wide) and a Scalar Engine handling complex floating-point operators. Compute engines are coupled to a multi-ported SRAM scratchpad and to eight custom DMA-engines, delivering the 102 GB/s bandwidth. The hardware logic resources of the NPU design are 321K LUTs, 458K FFs, and 2232 DSPs, and is clocked at 250 MHz with extensive clock gating logic.

## IV. Sensitivity Analysis (KV-Caching Context)

This section explores the impact of certain inference-time hyperparameters on performance. The goal is to select optimal settings for each backend to ensure fair and peak-performance comparisons during the subsequent evaluation. We will measure performance as the number of tokens per second obtained in inference, after prompt processing has been finished.

KV caching is a widely adopted technique that reduces the computational complexity of the self-attention mechanism from quadratic to linear with respect to sequence length. It achieves this by avoiding the re-computation of key and value tensors for previously processed tokens. This technique represents a trade-off between computation and memory: while it reduces compute overhead, the cached tensors occupy memory resources competing with model weights for bandwidth and capacity, which can affect overall performance. The size of the KV cache grows proportionally with the total sequence length (i.e., prompt length plus generated tokens), with an upper bound defined by the model's context window. The increase in KV cache size impacts multiple components of the self-attention mechanism: it enlarges the overall memory footprint, increases the computational cost of matrix multiplications, and amplifies the complexity of softmax operations. Together, these factors contribute to higher memory requirements and reduced efficiency as sequence length grows. To evaluate the impact of KV cache size on inference performance, we conducted a design space exploration, measuring the degradation in throughput (tokens per second) as the number of tokens generated increases. Figure 2, Figure 3 and Figure 4 illustrate the results across CPU, GPU ad NPU backends respectively, with performance values normalized to the shortest sequence length, corresponding to the smallest KV cache.

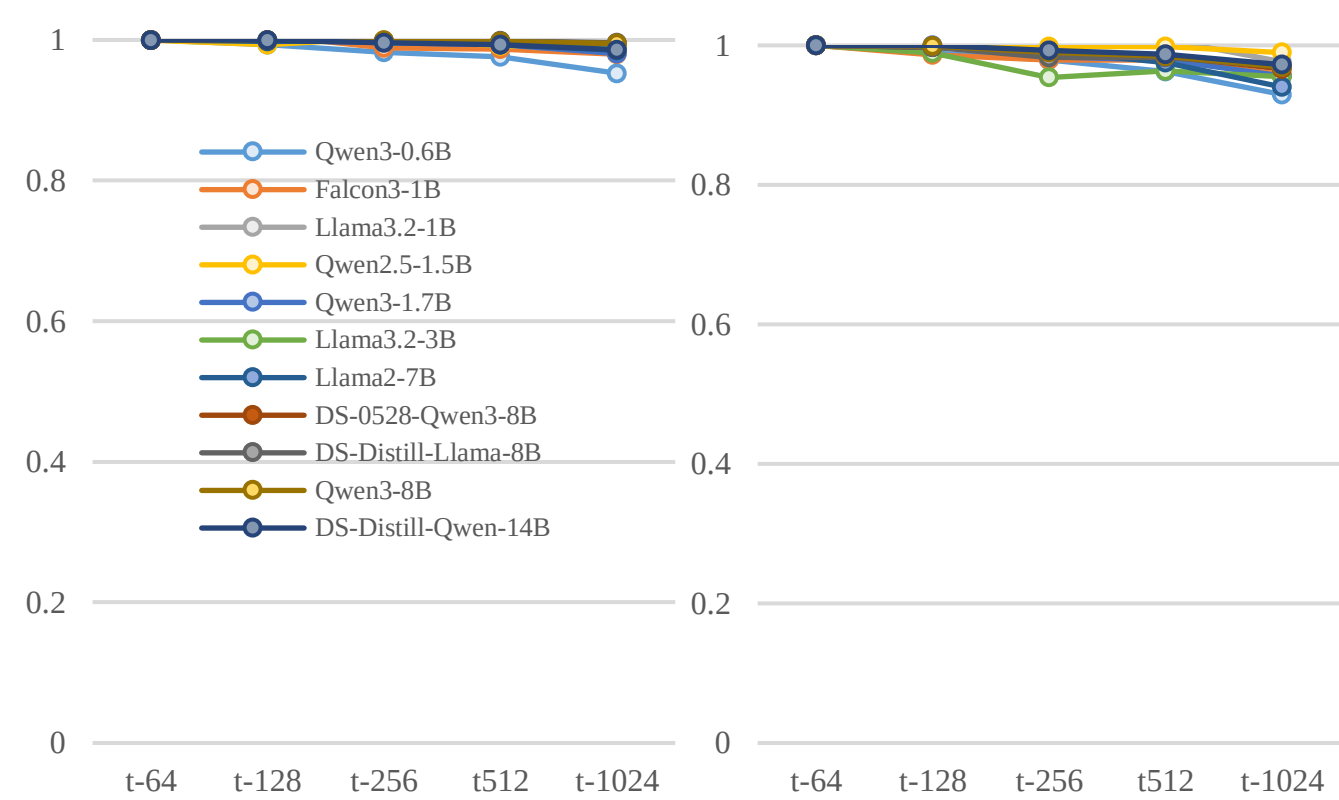

Figure 2. Performance slowdown in CPU backend as KV-cache size increases through larger token generation for F16 (left) and Q4K (right) data types.

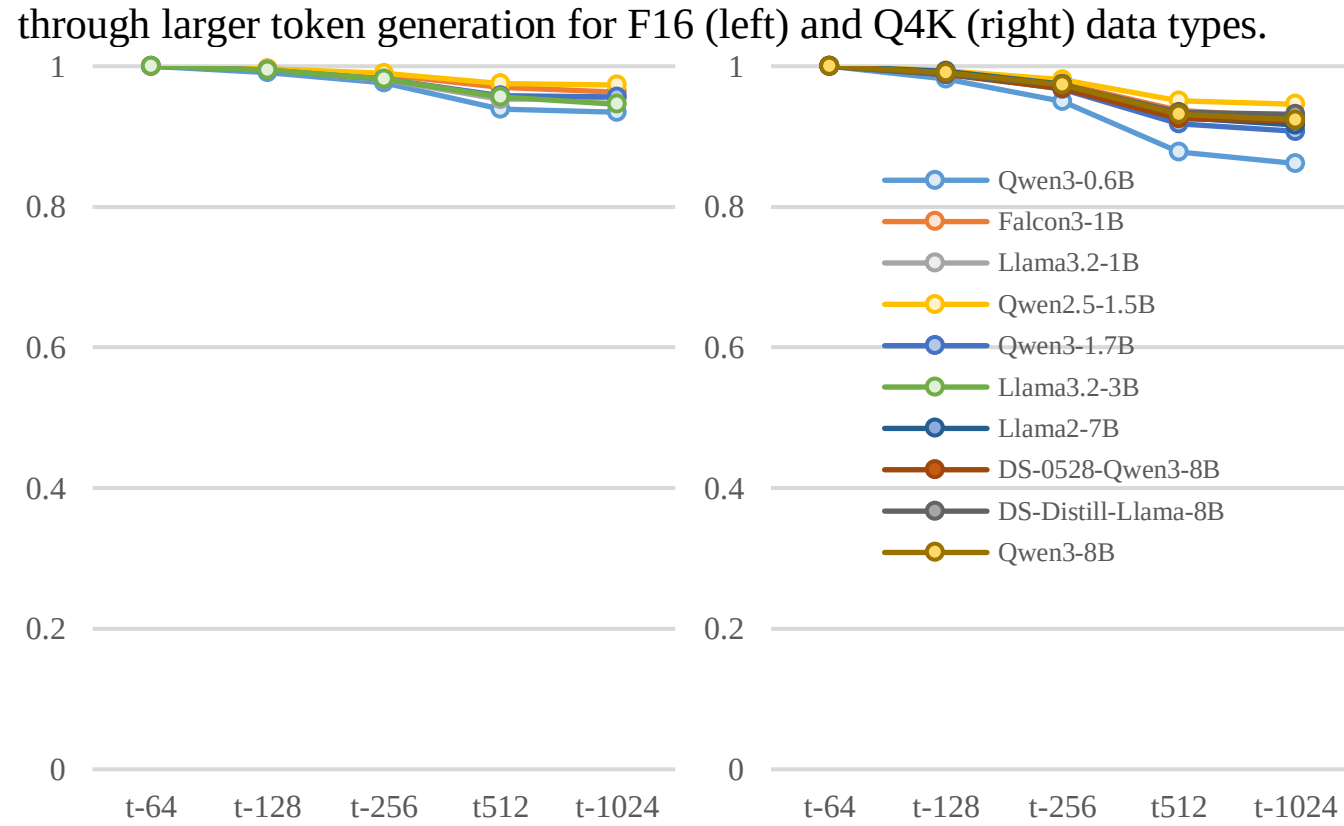

Figure 3. Performance slowdown in GPU backend as KV-cache size increases through larger token generation for F16 (left) and Q4K (right) data types.

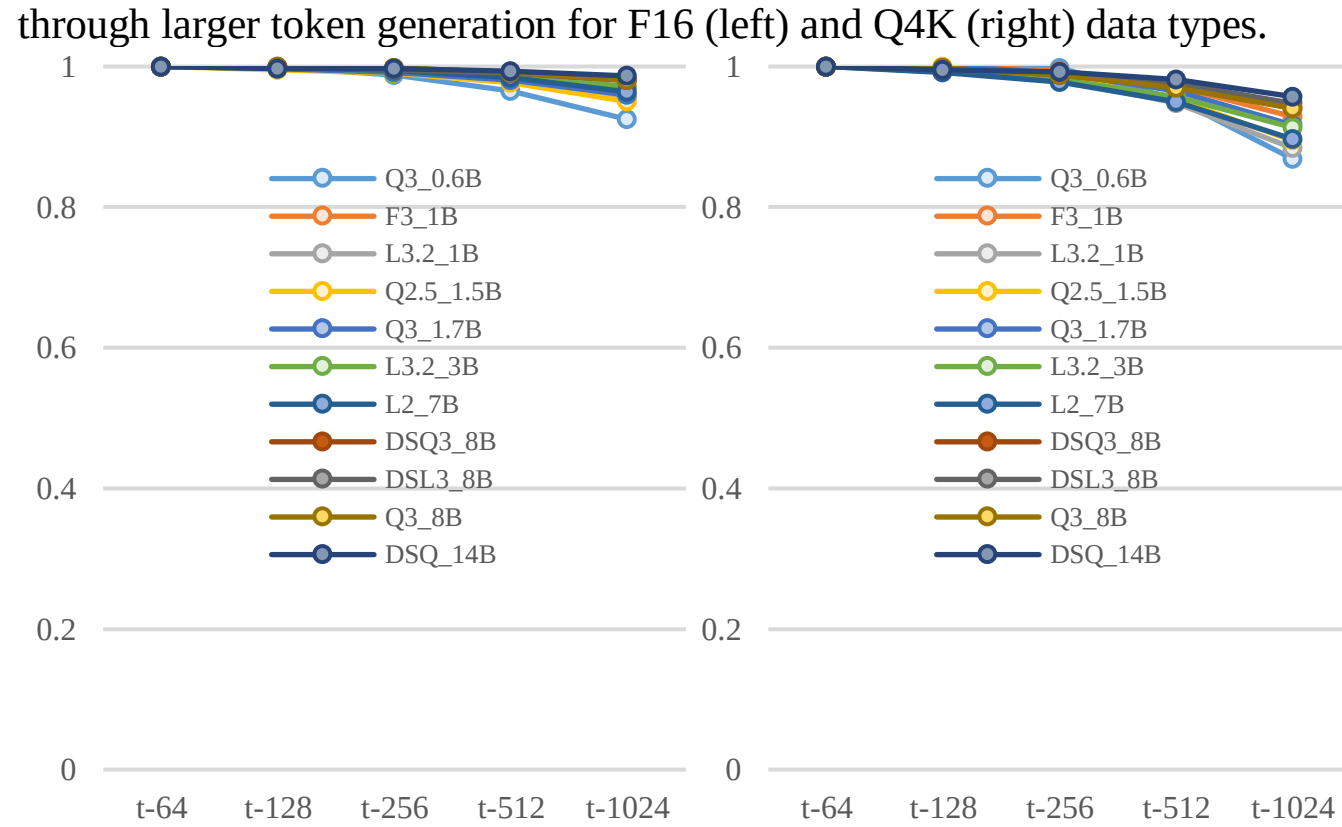

Figure 4. Performance slowdown in NPU backend as KV-cache size increases through larger token generation for F16 (left) and Q4K (right) data types.

Overall, performance degradation with increasing sequence length is limited, although it is strongly influenced by weight precision and model size. F16 models exhibit only modest reductions in throughput, whereas quantized models show a nearly constant loss, particularly on GPUs and NPUs. The behavior observed in the GPU and NPU may be explained by the increased cost of softmax operations, due to the limited availability of compute engines for this type of computation.

With respect to model size, smaller models tend to experience larger slowdowns, mainly because distributing computation across multiple units yields diminishing returns at that scale. In contrast, larger models show limited performance degradation, typically not exceeding 10%. While not strictly required in most cases, subsequent experiments in this study adopt a sequence length of 1024 tokens. This configuration provides a representative context length that makes the impact on performance observable, while keeping the KV-cache size manageable, enabling the evaluation of larger models on memory-constrained backends, such as the NVIDIA Jetson platform.

## V. EVALUATION

This section presents the results of the evaluation process, where each small language model (SLM) was executed on all hardware backends under test. Results in each figure are ordered by model size, with the smallest model displayed on the left. Each value corresponds to the average of 10 inference runs, with error bars indicating standard deviation. It is important to note for all the results shown in this section that the largest models (those with 7 billion parameters or more in F16 models and 14 billion or more in quantized models) have a working set greater than the available memory in two of the backends (NVIDIA Jetson GPU and ARM). Thus, these models could not be evaluated, resulting in a gap in the results graphs for GPU and ARM.

### *A. Raw Performance*

We begin our analysis by measuring the raw inference performance of each model, expressed in tokens generated per second. This metric reflects the system's ability to complete language generation tasks under ideal conditions. Each precision format (e.g., F16, Q4K) is evaluated independently. The corresponding performance results are grouped and visualized in Figure 5, with one graph per data type. In this graph, the y-axis represents the average tokens per second obtained for a 1024-token prediction.

Several observations can be drawn from these results. First, F16 among CPU-based backends, both the desktop processor and the ARM architecture achieve comparable performance, consistently below that of specialized accelerators. The performance of the ARM CPU could be surprising, especially in the F16 models, as it outperforms the X86 backend, although this difference will be explained later. Regarding non-general-purpose backends, although GPUs are typically considered the standard for this type of workload, the results indicate that a well-optimized NPU design can achieve higher throughput in both F16 and Q4 precision.

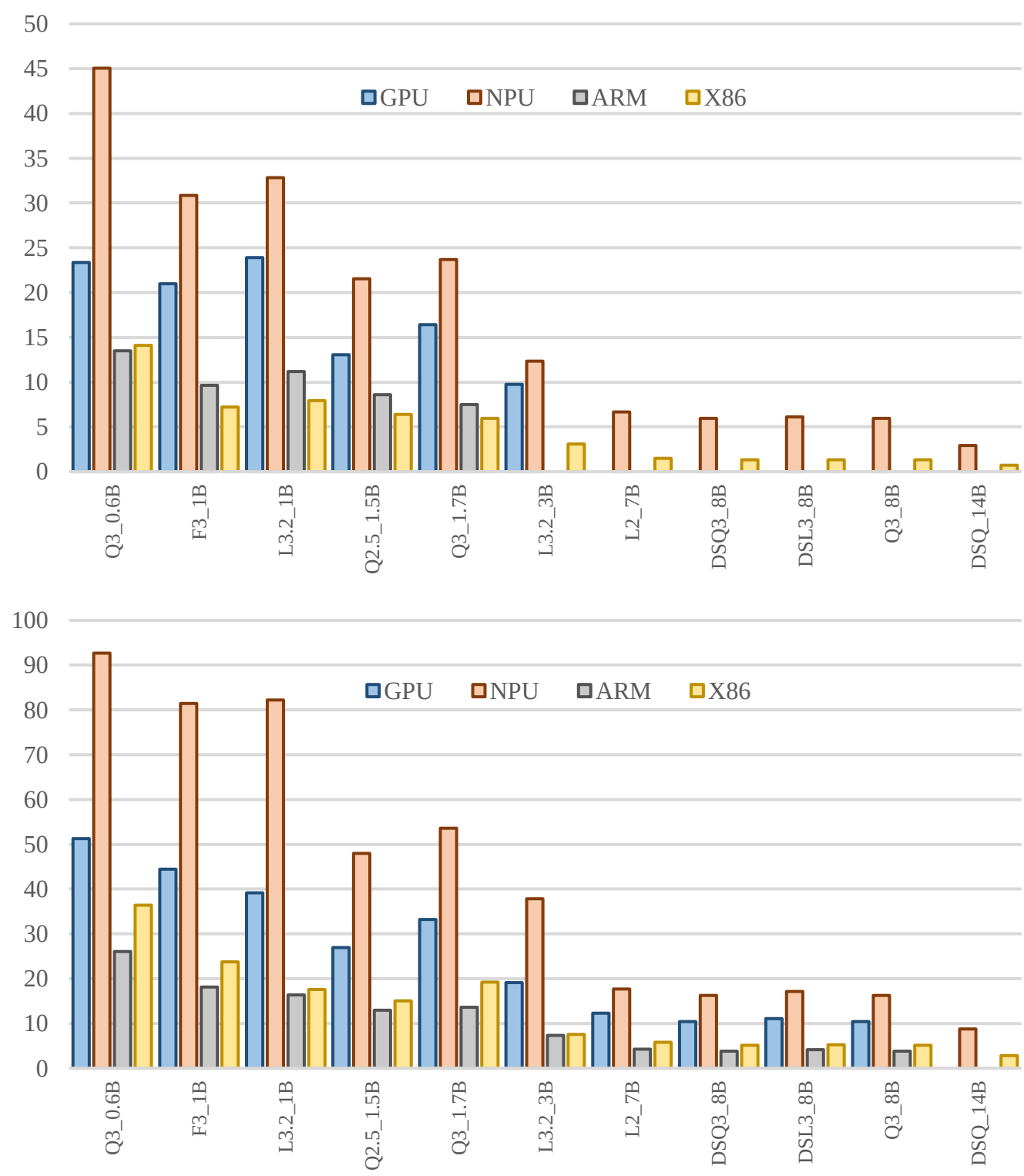

Figure 5. Performance obtained (tokens per second) for all the models evaluated running in the alternative backends. F16 (top) and Q4 (bottom) weights. Higher is better.

To facilitate comparison, we redraw in Figure 6 performance results, normalizing values obtained to the reference backend (GPU), allowing differences across backends to be observed independently of the performance variations introduced by model size. Since some of the models (F16 with 7B parameters and above) cannot be executed on the reference backend (GPU), their results are omitted from the figure. As shown in Figure 6, CPUs consistently achieve around 50% of the GPU throughput, with the ARM processor performing better at F16 and the x86 architecture in Q4K. On the other hand, the RaiderChip specialized NPU emerges as the best-performing platform, delivering more than 50% and 70% improvement over the GPU for F16 and Q4K precision formats respectively.

Despite the fact that some preliminary conclusions can be extracted from raw tokens-per-second results, direct performance comparisons across backends can be misleading due to architectural differences (e.g., CPU vs GPU vs heterogeneous SoC). To enable more meaningful comparisons, the rest of the paper tries to evaluate both hardware and energy efficiency using normalized metrics.

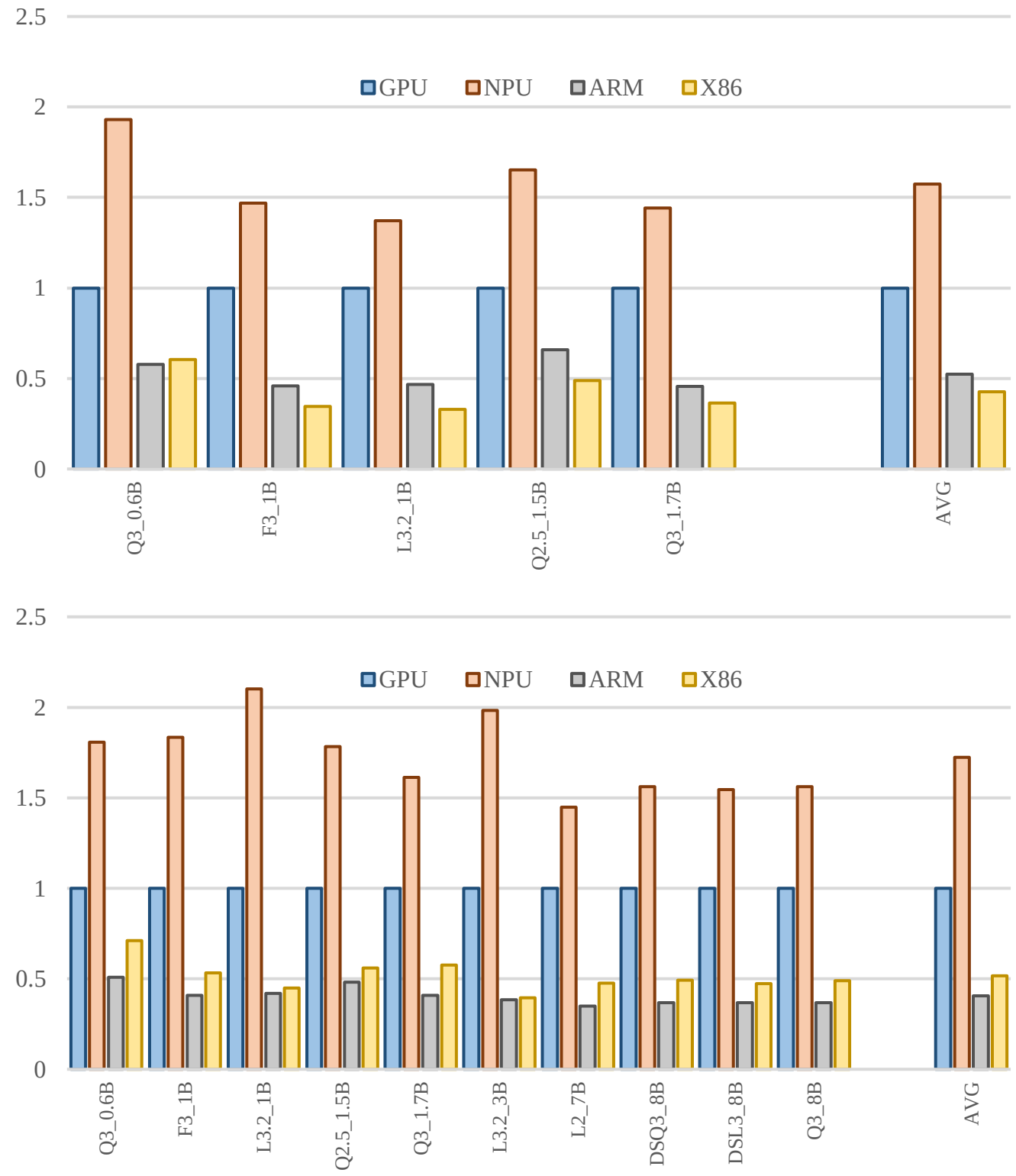

Figure 6. Normalized performance to reference values (GPU). F16 (top) and Q4 (bottom) weights. Higher is better.

### B. Hardware Efficiency

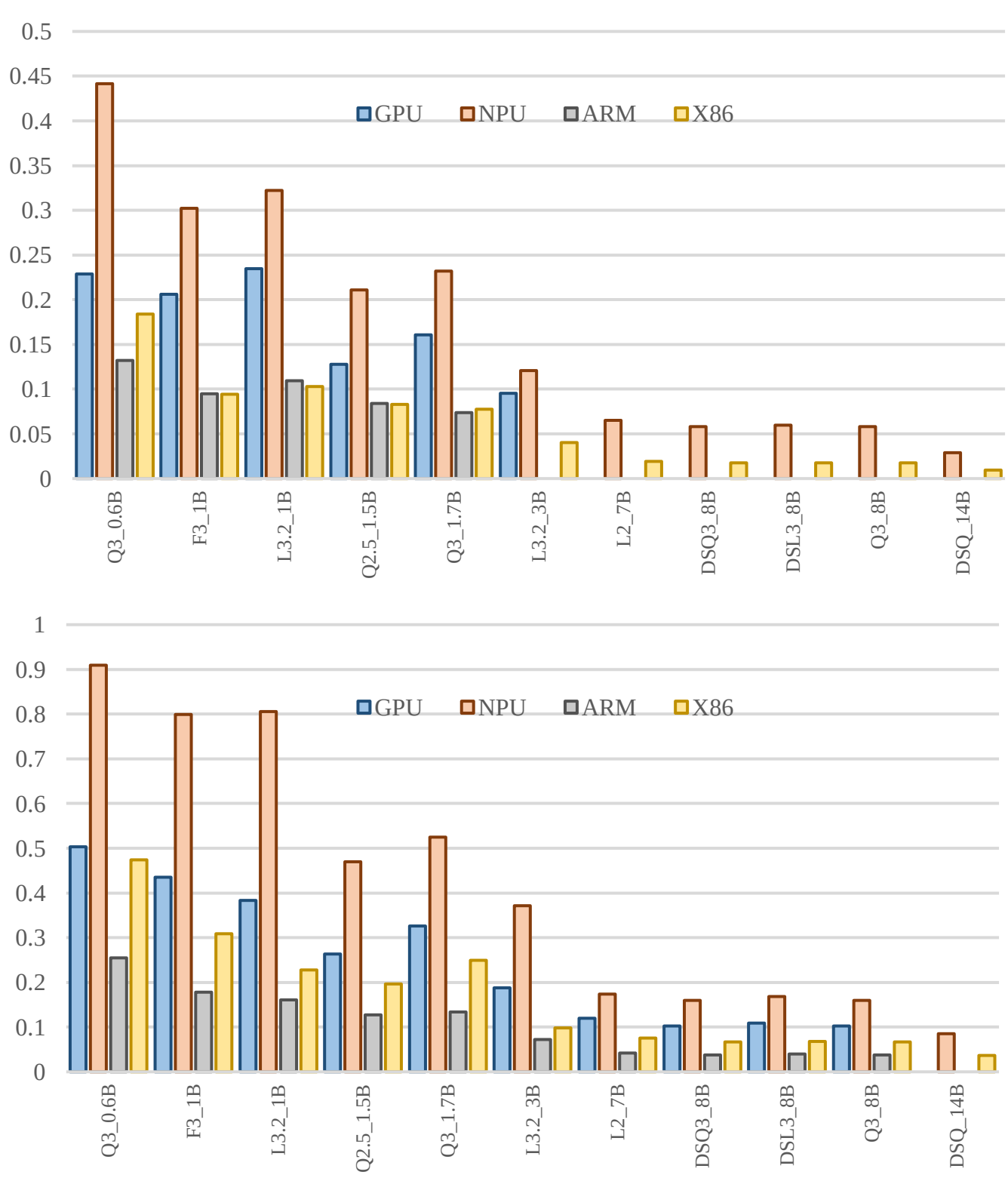

Figure 7. Performance obtained (tokens per second) normalized to the peak available bandwidth for each backend. F16 (top) and Q4 (bottom) weights. Higher is better.

Given that memory bandwidth is a primary bottleneck in transformer inference workloads [25], we normalize each model's throughput by the peak theoretical memory bandwidth of its respective backend. This provides insight into how effectively each architecture utilizes its available bandwidth. Normalized results are shown in Figure 7, allowing us to compare relative efficiency independent of absolute bandwidth capacity.

The results indicate that this metric consolidates the performance of all evaluated architectures. Notably, in the Q4 models, X86 shows improved performance, bringing it closer to the GPU. Since the bandwidth of the GPU and NPU was intentionally similar, their relative performance remains unchanged. Although the X86 backend gains some normalized performance due to its lower memory bandwidth, outperforming the ARM CPU, the specialized platforms, particularly the NPU, retain a clear performance advantage under this metric.

### C. Energy Efficiency

In edge computing environments, energy efficiency is often as critical as performance. Beyond latency, the power consumption per generated token becomes a key metric for system-level optimization. Backends used in this evaluation differ significantly in thermal design power (TDP) and architectural design.

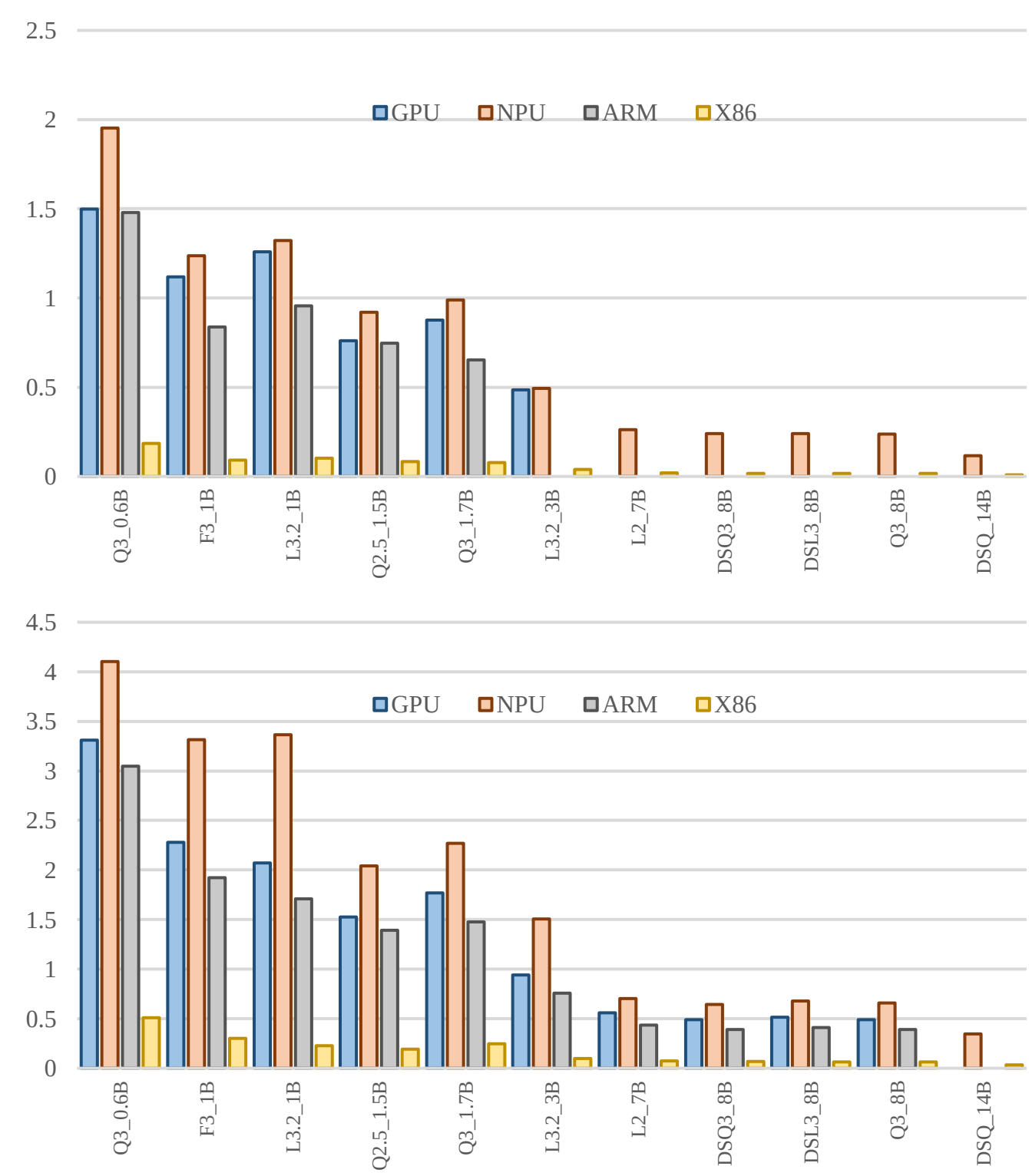

Figure 8. Energy efficiency measured as tokens generated per joule consumed for each backend. F16 (top) and Q4 (bottom) weights. Higher is better.

To assess energy usage, we measured average power consumed during inference and calculated the tokens generated per unit of energy. Power consumption varies substantially across backends, ranging from ~10 W for the ARM processor to ~75 W for the X86, with intermediate values of ~20 W for the GPU and ~24 W for the NPU. Based on these measurements, energy efficiency was evaluated in terms of tokens predicted per joule, as shown in Figure 8.

When considering total system consumption (measured at the system input), the x86 server-grade CPUs are clearly disadvantaged, confirming their limited suitability for edge scenarios. On the other hand, the ARM CPU, owing to its low power demand, achieves efficiency levels close to those of the GPU.

For the NPU, the advantage observed in normalized performance is reduced due to the high power consumption of the FPGA where it is implemented. Even so, in all the models evaluated, although reduced, the NPU still exhibits an efficiency advantage due to its raw performance.

In evaluating computing systems, considering only performance or only energy consumption often provides an incomplete picture. Pure performance metrics, such as execution time or throughput, reward faster designs but ignore the potentially prohibitive energy costs (in this scenario we could have chosen a high-end GPU or CPU, unsuitable for edge computing). Conversely, energy-focused measures highlight efficiency but may favor overly conservative designs that sacrifice speed (as can be seen in this paper with the ARM platform).

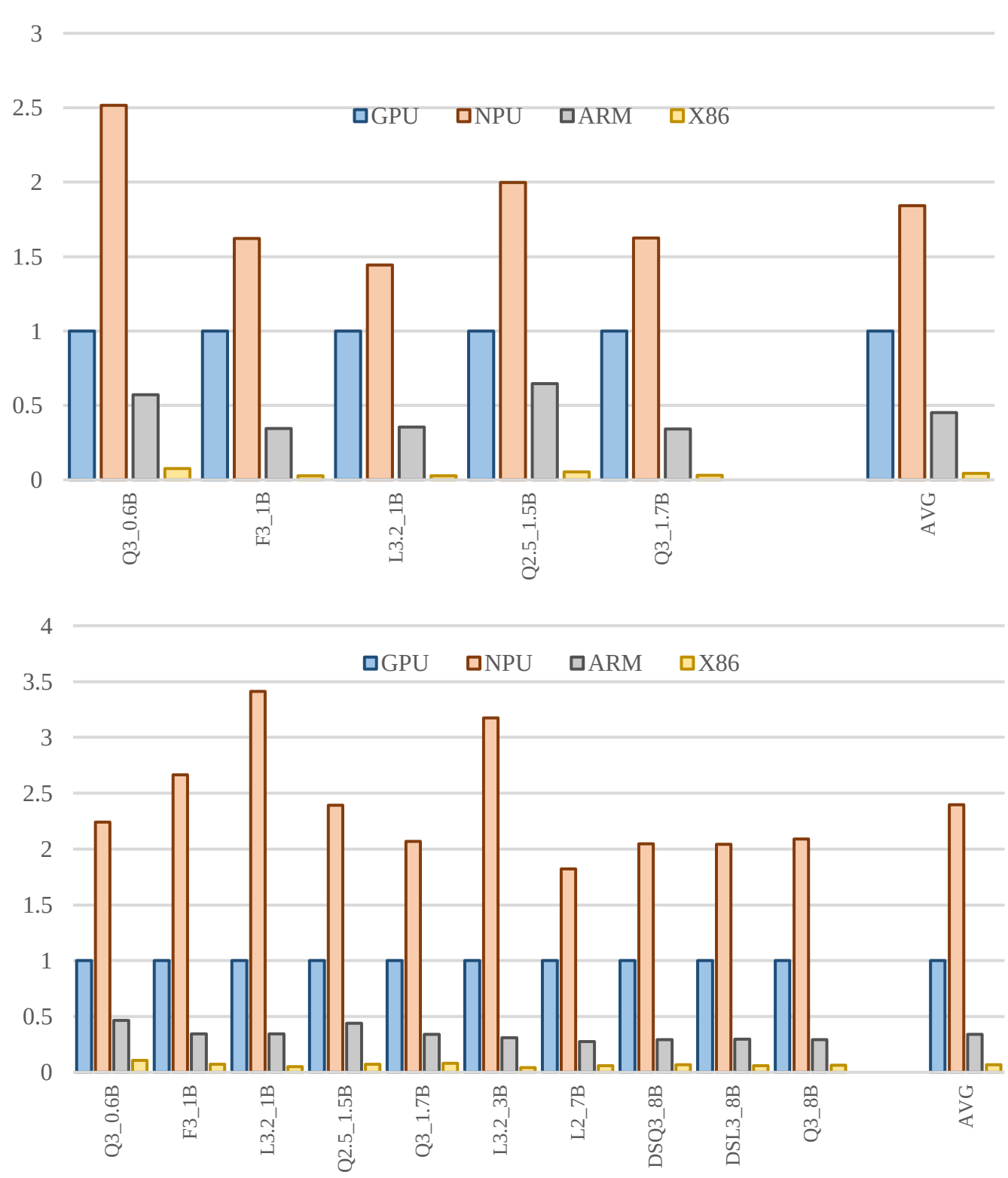

Figure 9. 1/EDP (Energy Delay Product) normalized to the GPU backend results. F16 (top) and Q4 (bottom) weights. Higher is better.

To balance these competing objectives, the energy–delay product (EDP) has become a widely adopted metric in architecture comparisons [26], as it jointly captures both energy efficiency and performance by penalizing solutions that excel in one dimension while neglecting the other. By incorporating the product of energy consumption and execution delay, EDP reflects the practical trade-off between minimizing power usage and maintaining responsiveness. In this work, our last experiment depicted in Figure 9 uses the inverse of the EDP so that higher values consistently indicate better outcomes, aligning the interpretation of this metric with the rest of our results.

The Figure reveals distinct behaviors depending on the backend. First, the ARM CPU achieve mid-level EDP values, considerably higher than those of x86 CPU but clearly lower than those of the specialized backends. In the case of ARM, limited performance is partially offset by low energy consumption, whereas the x86 system combines higher power usage with comparable performance loss, resulting in poor efficiency in this metric. Second, both GPU and NPU backends more than double the ARM's performance in EDP, benefiting from the combination of mid power consumption and strong raw performance. Among these, the NPU ad hoc design consistently delivers the best results, with an 84% average improvement over the GPU when executing F16 models and 140% improvement on average in this metric for Q4 models.

## VI. Conclusions

This study has presented a comprehensive evaluation of CPUs, GPUs, and NPUs as backends for Small Language Model (SLM) inference in edge environments. As demonstrated by the results, while raw performance results allow some preliminary conclusions, a further analysis should be provided when comparing such different platforms.

When examining normalized efficiency metrics, memory bandwidth emerges as the dominant factor influencing inference throughput. In this work, the results for the x86 CPU with lower memory bandwidth become more significant. However, since the evaluated backends were intentionally selected to have broadly similar bandwidth characteristics, the impact of this metric is reduced. Nevertheless, its inclusion remains essential to ensure fairness and completeness in any study of this kind.

Energy consumption is another key factor for edge deployment. In this regard, ARM processors stand out as low-power alternatives, offering moderate efficiency primarily due to their reduced energy usage. Among specialized backends, the NPU achieves better results than the GPU, thanks to a constrained energy consumption combined with a superior token generation throughput. In contrast, the x86 platform lags significantly behind the other backends because of its higher power demand. The surprising energy efficiency of the ARM platform further motivates the use of composite metrics such as the Energy–Delay Product (EDP), which account for energy efficiency and performance while penalizing cases where improvements are achieved in only one of these dimensions. The EDP analysis confirms that specialized backends provide a much better balance between performance and cost (energy consumption in this case) when compared to general-purpose platforms. Focusing on specialized hardware, RaiderChip's NPU clearly surpasses NVIDIA GPU (in every metric), reaching an 84% and 140% EDP improvement for the F16 and Q4K models respectively.

## Acknowledgment

This work was supported by the Spanish Government (Ministerio de Ciencia, Innovación y Universidades / Agencia Estatal de Investigación) under grant PID2022-139664NB-I00.